\documentclass[singlespacing]{elsart}

\usepackage{graphicx}

\usepackage{amssymb}
\journal{Chemical Physics Letters}
\begin{document}
\begin{frontmatter}

\title{Model for vibrationally enhanced tunneling of proton transfer in hydrogen bond}
\author{ A.E. Sitnitsky},
\ead{sitnitsky@kibb.knc.ru}
\address{Kazan Institute of Biochemistry and Biophysics, FRC Kazan Scientific Center of RAS, P.O.B. 30,
420111, Russian Federation. e-mail: sitnitsky@kibb.knc.ru}
\begin{abstract}
Theoretical analysis of the effect of an external vibration on proton transfer (PT) in a hydrogen bond (HB) is carried out. It is based on the two-dimensional Schr\"odinger equation with trigonometric double-well potential. Its solution obtained within the framework of the standard adiabatic approximation is available. An analytic formula is derived that provides the calculation of PT rate with the help of elements implemented in {\sl {Mathematica}}. We exemplify the general theory by calculating PT rate constant for the intermolecular HB in the Zundel ion ${\rm{H_5O_2^{+}}}$ (oxonium hydrate). This object enables one to explore a wide range of the HB lengths. Below some critical value of the frequency of the external vibration the calculated PT rate yields extremely rich resonant behavior (multiple manifestations of bell-shaped peaks). It takes place at symmetric coupling of the external vibration to the proton coordinate. This phenomenon is absent for anti-symmetric and squeezed mode couplings.
\end{abstract}
\begin{keyword}
Schr\"odinger equation, double-well potential, quantum tunneling, spheroidal function, Zundel ion.
\end{keyword}
\end{frontmatter}
\section{Introduction}
Proton transfer (PT) in hydrogen bonds (HB) is one of the main processes in the reaction rate theory. It takes place in the most important biological molecules such as proteins (participating in some enzymatic reactions) and DNA (arguably participating in the occurrence of mutations). In particular the phenomenon of vibrationally enhanced (or assisted or promoted) tunneling at PT (i.e., resonant acceleration of the process by a coupled oscillation in some frequency range) \cite{Sok92}, \cite{Ham95} is of great interest especially in regard of its possible role in a mechanism for enzymatic hydrogen transfer [3-6]. Within the context of enzyme catalysis it is a specific case of the more general trend named "rate-promoting vibration" [6-11]. There are several cases in which a vibration can be coupled to the proton coordinate in HB. First of all there is the heavy atoms stretching mode which is an intrinsic degree of freedom in HB. It is thoroughly studied theoretically since the pioneer articles \cite{Mey87}, \cite{Mey90}, \cite{Sok92} dealing with vibrationally promoted PT in solids. Unfortunately it is an internal vibration and its fixed frequency is not an experimentally controllable parameter. Then there are external vibrations exerted on HB and provided either by protein scaffold for HB in enzymes or by some means from the researcher's toolkit for HB in model compounds. One of the most efficient ways for such purpose is the usage of the IR electromagnetic field of an optical cavity. The phenomenon of resonant activation (or, in contrast, suppression) of reaction rates is widely discussed for modifying chemical kinetics by optical cavities (for recent articles in this field which is sometimes called vibrational polariton chemistry see, e.g., [14-16] and refs. therein). The resonance, i.e., maximal cavity induced enhancement of the reaction rate under the vibrational resonance condition is produced in this case by mixing the electromagnetic field with quantum states of molecular systems. The cavity is equivalent to a harmonic oscillator of a given frequency coupled to the molecular system. The Hamiltonian of the molecule degree of freedom coupled to the field oscillator in the electric dipole approximation of light-matter interaction has the same structure as those used for PT coupled to the heavy atoms stretching mode in HB. In this regard constructing reliable theoretical models of PT which take into account the possibility of varying the frequency of the external vibration exerted on HB is a long-standing problem for the reaction rate theory and seems to be of interest for perspectives of various application.

A proton in HB is known to be sufficiently light to exhibit full-fledged quantum behavior leading to tunneling effect, energy levels splitting, etc (see, e.g.,
[12,13,17-28] and refs. therein). Physical models of PT based on simplified Hamiltonians take their peculiar place in the enormous amount of literature on HB including also  {\it ab initio} calculations by the methods of quantum chemistry, DFT, their combination with molecular dynamics simulations (considering nuclei as classical Newtonian particles), QM/MM, chemical physics approaches within the framework of modern trends in TST, quantum-classical Liouville dynamics, etc. In physical models of PT the reaction coordinate is singled out and studied separately from the environment for which various approximations are assumed. The problem of PT rate estimate is inevitably reduced to a one-dimensional path on the potential energy surface (PES) and as a result to a one-dimensional cross-section of PES for HB which usually has the form of a double-well potential (DWP). Quantum mechanical models of PT are motivated by the necessity to take into account other (than the reaction coordinate) internal dynamic modes, e.g., the heavy atoms stretching mode and to account for vibrationally and/or thermally assisted tunneling. The modern physical approach to taking into account dissipative effects at tunneling is based on the Lindblad master equation (describing the dynamics of Markovian open quantum systems) for the time evolution of the density matrix and Caldeira-Leggett model of the thermal bath. For PT such scheme was initiated in [12,13] and by now it has been thoroughly studied within the framework of the general context for the reaction rate theory (see, e.g., \cite{God15} and refs. therein). The problem of the rate-promoting vibration for PT is considered with the help of this theory in \cite{Shi11}. The authors obtained the desired increase of the PT rate at adding the vibrational mode. However they came to a conclusion that the lower its frequency the stronger the enhancement of PT rate. Our aim is to find out the conceptual possibility of resonant activation (bell-shaped peaks) with frequency. In the present article we avoid the complications of the above mentioned theory (ensuing from the necessity to deal with numerous evolution equations for the density matrix elements) which seem to be unimportant for our aims. For calculating PT rate we make use of the Weiner's theory [18,19]. Our model of HB is maximally simple and constructed in an ad hoc manner for studying the effect of PT resonant activation. It corresponds to HB in a gas phase and does not touch upon the effects of environment taking place in solution. It deals only with two salient degrees of freedom, i.e., the proton coordinate and that of an oscillator (e.g., the heavy atoms stretching mode or an external vibration) with symmetric coupling between them. We treat both degrees of freedom quantum-mechanically by solving the corresponding two-dimensional Schr\"odinger equation (SE). We make use of literature data for the one-dimensional cross-section of PES from quantum-chemical calculations and model it by a suitable phenomenological DWP. For the case of the heavy atoms stretching mode we use literature data of IR-spectroscopy for HB to determine its frequency and the strength of proton coordinate coupling to it.

The calculation of PT rate in HB requires the knowledge of the energy levels which are the eigenvalues of the corresponding SE with DWP. PT is a typical example of a quantum particle in DWP which is an omnipresent problem in physics and chemistry [23,26,29-41]. Most DWPs used for the analysis of HB and composed of polynomials, exponentials (e.g., the double Morse potential) or their combinations are amenable only to numerical solutions (even in one-dimensional case let alone its two-dimensional generalization) or approximate analytic approaches like the quasi-classical (WKB) method. This restriction was inevitable until 2010-s because of the lack of a convenient DWP for which SE would have an exact analytic solution (see \cite{Jel12} and refs. therein). Since then a number of exactly solvable DWPs suitable for chemical problems (taking infinite values at the boundaries of the spatial variable interval) appeared. For them analytic solutions of SE are feasible via the confluent Heun's function [23], [31-38] or the spheroidal function [39,40]. The latter is a well-studied special function of mathematical physics \cite{Kom76} implemented in {\sl {Mathematica}}. The case which is amenable to the treatment by both functions [23,31,39] makes use of the so-called trigonometric DWP (TDWP). In the previous years TDWP was applied to numerous objects [23,24,25,31,39,40,43,44]. The aim of the present article is to show that TDWP enables one to construct an analytically tractable model for PT resonant activation in HB. We exemplify the general theory by the analysis of PT rate for intermolecular HB in the Zundel ion ${\rm{H_5O_2^{+}}}$ (oxonium hydrate ${\rm{H_2O\cdot\cdot\cdot H \cdot\cdot\cdot OH_2}}$ in which the proton is equally shared between two water molecules). For the Zundel ion the detailed data of IR spectroscopy [20-22] along with the quantum chemical {\it ab initio} calculations [45,46] are available. As a result the Zundel ion suits well for the purpose of demonstrating the capability of our approach to the calculation of PT rate in HB. For the Zundel ion the distance between the oxygen atoms $R_{OO}$ is not a fixed and predetermined value but can be varied in a wide range. In the present article the case $R_{OO}=3.0\ A$ is chosen because it provides sufficiently high barrier to exclude the contribution of the over-barrier transition into PT rate constant even at high temperature. This choice is in accord with the aim of the article to study the effect of an external vibration on the tunneling contribution into the rate constant.

The paper is organized as follows. In the preliminary Sec.2 we remind some results of the Weiner's theory in the form suitable for our analysis. In Sec.3 we briefly summarize the results of \cite{Sit20} which are necessary for the calculation of PT rate in HB. In Sec.4 we derive the expression for the PT rate constant. In Sec.5 the results are discussed and the conclusions are summarized. In Appendix some technical information is presented.
\section{Weiner's theory}
In the Weiner's theory [18,19] the proton position is described by the stationary one-dimensional SE (which we write in the dimensionless form) with symmetric DWP $U(x)$ which has the solutions for the energy levels $\epsilon_q$ and the corresponding wave functions $\psi_q (x)$
\begin{equation}
\label{eq1}  \psi''_q(x)+\left[\epsilon_q-U(x)\right]\psi_q(x)=0
\end{equation}
The rate constant consists of the contribution from the tunneling process and that from the over-barrier transition. Concerning the former the Weiner's theory deals with two important values. The first one is the probability flux to the right of particles in the left well when the particle is in the q-th state $J_q$. The second one is the quantum transmission coefficient, i.e., the fraction of those right-moving particles which are transmitted to the right well $\mid T_q \mid^2$. According to [18,19] the reaction rate constant is a result of Boltzmann averaging of the product $J_q\mid T_q \mid^2$ calculated over the doublets
\begin{equation}
\label{eq2}  k=\left[\sum_{q=0}^{\infty}e^{-\beta \epsilon_q}\right]^{-1} \left\{\sum_{n=0}^N e^{-\beta \epsilon_{2n}} J_{2n} \mid T_{2n} \mid^2+\sum_{m=2N+2}^{\infty}e^{-\beta \epsilon_m}\right\}
\end{equation}
where $n=0,1,2,..., N\ $, $\epsilon_{2n}$ is the energy for the level $2n$ described by the wave function $\psi_{2n} (x)$. In the Weiner's theory the quantum transmission coefficient is calculated for the doublets which are counted by the even energy levels. For this reason $n$ is fixed to be even in the first sum in the curly brackets (see the text below the formula (3.1) in Sec.III of [19] the formulas from which are used in the present article). The first sum in the curly brackets corresponds to the contribution due to the tunneling process in the reaction rate. It is over the energy levels below the barrier top for which the notions of $ J_{2n}$ and $ \mid T_{2n} \mid^2$ have sense. In (\ref{eq2}) it is suggested by Weiner that the quantum transmission coefficient of the lower level in the doublet is determined by the splitting of the energy levels in it. Thus $N+1$ is the number of doublets below the barrier top and $\epsilon_{2N}$ is the lower energy level in the last doublet in this region. As a result only the sum over doublets (i.e., even levels $q=2n$) is left. The second sum in the curly brackets corresponds to the over-barrier transition and $\epsilon_{2N+2}$ is the the first energy level above the barrier top. The Weiner's theory is based on the quasi-classical approximation of the solution of SE \cite{Wei78a}
\begin{equation}
\label{eq3}  \psi_{2n} (x)=\frac{B_{2n}}{\sqrt {P_{2n}(x)}}\cos\left(\int_0^x d\xi\ P_{2n}(\xi) + S_{2n}\right)
\end{equation}
for $x\geq 0$. Taking into account that for even energy levels the wave function is symmetric ($\psi'_{2n} (0)$) one obtains that
\begin{equation}
\label{eq4}  \tan S_{2n} =-\frac{P'_{2n}(0)}{2P^2_{2n}(0)}
\end{equation}
The function $P_q(x)$ satisfies the so-called Milne equation
\begin{equation}
\label{eq5}  P_q^2+U(x)+\frac{1}{2}\left[\frac{P_q''}{P_q}-\frac{3}{2}\frac{\left(P_q'\right)^2}{P_q^2}\right]=\epsilon_q
\end{equation}
The expression for  $\mid T_{2n}\mid^2$ follows from (3.5) of \cite{Wei78a}
\begin{equation}
\label{eq6}  \mid T_{2n} \mid^2=\frac{\psi_{2n}^2 (0)P_{2n}(0)}{B_{2n}^2}
\end{equation}
The expression for $J_{2n}$ is given by (2.14) of \cite{Wei78a}
\begin{equation}
\label{eq7}  J_{2n}=\frac{B_{2n}^2}{2}
\end{equation}
In the particular case $P'_{2n}(0)=0$ (which will be pertinent in our further consideration) it follows from (\ref{eq4}) that
\begin{equation}
\label{eq8} S_{2n}=0
\end{equation}
Substitution of the results into (\ref{eq2}) yields
\begin{equation}
\label{eq9}   k=\left[\sum_{q=0}^{\infty}e^{-\beta \epsilon_q}\right]^{-1} \left\{\frac{1}{2}\sum_{n=0}^N e^{-\beta \epsilon_{2n}} B_{2n}^2+\sum_{m=2N+2}^{\infty}e^{-\beta \epsilon_m}\right\}
\end{equation}
It is worthy to note that in the original Weiner's approach both $\epsilon_q$ and $\psi_q (x)$ are unknown and all efforts are directed to obtain formulas that do not contain values like $\psi_q(0)$ or $\psi_q' (0)$. In contrast for TDWP the exact solution of SE $\psi_q (x)$ as well as the corresponding energy levels  $\epsilon_q$ are available and we make use of the them. Also it should be stressed that the Weiner's theory is originally written for the infinite range of the space variable $-\infty < x < \infty$ with the requirement $\mid \psi_q (x) \mid \rightarrow 0$ at $x \rightarrow \pm \infty$. In our case of TDWP we have the requirement $\mid \psi_q (x) \mid \rightarrow 0$ at $x \rightarrow \pm \pi/2$. For this reason we apply the corresponding formulas to the case $-\pi/2 \leq x \leq \pi/2$. We consider SE (\ref{eq1}) in this range with the dimensionless form of the symmetric TDWP \cite{Sit18}
\begin{equation}
\label{eq10} U(x)=\left(m^2-\frac{1}{4}\right)\ \tan^2 x-p^2\sin^2 x
\end{equation}
Here $m$ is an integer number and $p$ is a real number. The two parameters of TDWP $m$ and $p$ are related to two main characteristics of the potential energy surface, i.e., the barrier height and the barrier width (see Appendix). The example of TDWP for intermolecular HB in the Zundel ion with $R_{OO}=3.0\ A$ distance between oxygen atoms is presented in Fig.1.
\begin{figure}
\begin{center}
\includegraphics* [width=\textwidth] {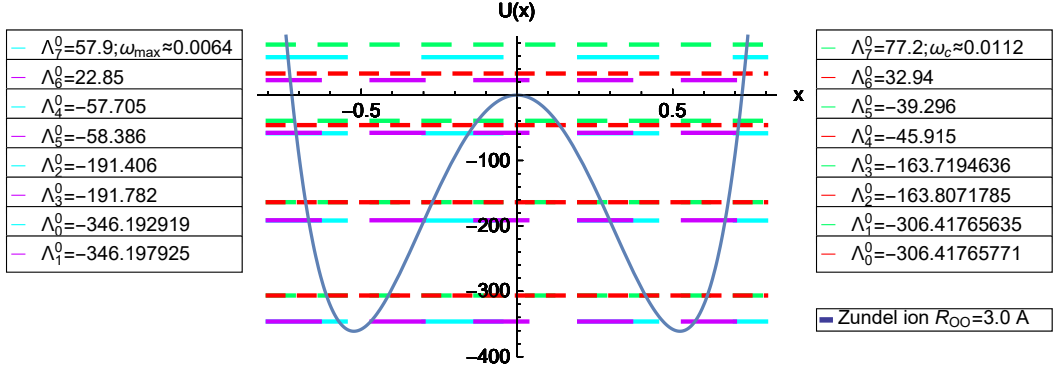}
\end{center}
\caption{The trigonometric double-well potential (\ref{eq10}) at the values of the parameters $m=57$; $p=76$. The parameters are chosen to describe the hydrogen bond in the Zundel ion ${\rm{H_5O_2^{+}}}$ (oxonium hydrate) for the case $R_{OO}=3.0\ \AA$ (they are extracted from the data of quantum chemistry \cite{Xu18}). Several energy levels are indicated for the coupling constant $\alpha=0.3$. They are given by (\ref{eq19}) and calculated at two frequencies of the external oscillator (for that of the main peak $\omega_{max} \approx 0.0064$ depicted by long dashes and for the critical frequency $\omega_c \approx 0.0112$ depicted by short dashes).} \label{Fig.1}
\end{figure}
For TDWP the exact solution of SE is available \cite{Sit18}
\begin{equation}
\label{eq11} \psi_q (x)=\cos^{1/2} x\ \bar S_{m(q+m)}\left(p;\sin x\right)
\end{equation}
$q=0,1,2,...$ and $\bar S_{m(q+m)}\left(p;s\right)$ is the normalized angular prolate spheroidal function \cite{Kom76}. It is implemented in {\sl {Mathematica}}
as $\rm{SpheroidalPS}[(q+m),m,ip,s]$ (note that the latter is a non-normalized one). The energy levels are
\begin{equation}
\label{eq12} \epsilon_q=\lambda_{m(q+m)}\left(p\right)+\frac{1}{2}-m^2-p^2
\end{equation}
Here $\lambda_{m(q+m)}\left(p\right)$ is the spectrum of eigenvalues for $\bar S_{m(q+m)}\left(p;s\right)$. It is implemented in {\sl {Mathematica}} as $\lambda_{m(q+m)}\left(p\right)\equiv \rm{SpheroidalEigenvalue}[(q+m),m,ip]$.\\
For TDWP the position of the right minimum is defined by the requirement
\begin{equation}
\label{eq13} \cos\ x_{min}=\left[\frac{m^2-1/4}{p^2}\right]^{1/4}
\end{equation}
\section{Solution of two-dimensional Schr\"odinger equation with trigonometric double-well potential}
For the two-dimensional SE the wave function is the function of the proton coordinate $x$ and that of the oscillator $z$. The interaction Hamiltonian for various types of the mode coupling can be schematically depicted by the form $\alpha f(x)g(z)$ where $\alpha$ is the coupling constant. For the symmetric mode coupling it is $\alpha z x^2$, for anti-symmetric and squeezed mode couplings it is  $\alpha z x$ and  $\alpha z^2 x^2$ respectively. In the present article we consider the case of the symmetric mode coupling (see a comment on other types of interaction in Sec.5). For TDWP it is more natural for the mathematical convenience to make for $x$ in the interaction term the transformation $x \rightarrow \sin x$ so that the coupling term is $\alpha z\sin^2 x$. In fact the external vibration always interacts with some function $f(x)$ of the proton coordinate $x$ (e.g., the with the dipole moment if the vibration is produced by an electro-magnetic field). The term $\alpha z x^2$ for the symmetric mode coupling means that only the linear approximation for the function $f(x)\approx x$ is taken into account. However the linear approximation can be valid within the interval of a sufficiently small $x$ only. In our opinion it is reasonable to go beyond the linear approximation, i.e., to make the replacing $x\longrightarrow \sin x$ at $-\pi/2 \leq x \leq \pi/2$. In the case of the dipole moment it deflects to slower growth than the linear one (see, e.g., Fig. 10.54 in \cite{Atk09}. The necessity to go beyond the framework of the linear approximation for HB in the Zundel ion was stressed in \cite{Ven01}. To achieve this goal we model such deflection by replacing the linear term by the trigonometric one $x\longrightarrow \sin x$ at $-\pi/2 \leq x \leq \pi/2$. Then the dimensionless form of the two-dimensional SE with the symmetric mode coupling and TDWP is \cite{Sit20}
\[
\Biggl\{\delta\frac{\partial ^2 }{\partial z^2}+\frac{\partial^2}{\partial x^2}+\Lambda-\left(m^2-\frac{1}{4}\right)\ \tan^2 x+p^2\sin^2 x-
\]
\begin{equation}
\label{eq14} \frac{\omega^2z^2}{2}-\alpha z\sin^2 x\Biggr \}\Phi(x,z)=0
\end{equation}
Here $\omega$ is the frequency of the oscillator coupled to the proton coordinate. The dimensionless variables and parameters are discussed in Appendix for the case when the oscillator is produced by the heavy atoms stretching mode in HB. The solution of (\ref{eq14}) in the adiabatic approximation corresponding to the $q$-th state of the particle in TDWP and the $j$-th state of the oscillator is \cite{Sit20}
\begin{equation}
\label{eq15} \Phi(x,q,z,j)\approx \varphi^q_j(z) \psi_q (x)
\end{equation}
Here the quantum number $q$ quantizes the states of the particle in TDWP and $\psi_q (x)$ is given by (\ref{eq11}). The quantum number $j$ in (\ref{eq15}) quantizes the excitation states of the oscillator
\begin{equation}
\label{eq16}
\varphi^q_j(z)\approx A \exp\left[-\frac{\omega}{2\sqrt{2\delta}}\left(z+\frac{c_q}{\omega^2}\right)^2\right]\frac{j!(-2)^j}{(2j)!} H_{2j}\left(\left(\frac{2\omega^2}{\delta}\right)^{1/4}\left(z+\frac{c_q}{\omega^2}\right)\right)
\end{equation}
$j=0,1,2,...\ $, $H_{n}(x)$ is the Hermit polynomial and A is a normalization constant. Making use of N2.20.16.6 from \cite{Pr03} we obtain
\begin{equation}
\label{eq17} A^{-2}=\left[\frac{j!(-2)^j}{(2j)!}\right]^2 2^{4j}\sqrt{\frac{\sqrt{2\delta}}{\omega}}\Gamma \left(2j+\frac{1}{2}\right)\ _2F_1\left(-2j,-2j,\frac{1}{2}-2j
;-\frac{1}{2}\right)
\end{equation}
where $\ _2F_1\left(a,b,c;x\right)$ is the hypergeometric function.
The coefficient $c_q$ is
\begin{equation}
\label{eq18} c_q=\alpha\int_{-1}^{1}d\eta\ \eta^2\ \left[\bar S_{m(q+m)}\left(p;\eta\right)\right]^2
\end{equation}
The energy levels corresponding to (\ref{eq15}) are \cite{Sit20}
\begin{equation}
\label{eq19} \Lambda^j_q\approx \lambda_{m(q+m)}\left(p\right)+\frac{1}{2}-m^2-p^2-\frac{\left(c_q\right)^2}{2\omega^2}+(4j+1)\omega \sqrt{\frac{\delta}{2}}
\end{equation}
\section{Proton transfer rate constant}
We introduce the dimensionless inverse temperature $\beta$ (for its expression via dimensional parameters of the model see Appendix). Further we restrict ourselves to the relatively high temperature range $200\ K \leq T \leq 400\ K$ ($0.0345 \leq \beta\leq 0.069$) in which the Boltzmann statistics is valid. Then the partition function is calculated with the help of the energy levels $\Lambda^k_q$ given by the formula  (\ref{eq19})
\[
 Z(\beta, \omega)=\sum_{q}\sum_{j=0}^{\infty}\exp\left[-\beta \Lambda^j_q(\omega)\right]=
\]
\begin{equation}
\label{eq20}
\sum_{j=0}^{\infty}e^{-\beta \left[(4j+1)\omega \sqrt{\frac{\delta}{2}}+\frac{1}{2}-m^2-p^2\right]}\sum_{q}e^{-\beta \left[\lambda_{m(q+m)}\left(p\right)-\frac{\left(c_q\right)^2}{2\omega^2}\right]}
\end{equation}
With the help of (\ref{eq16}) we calculate the average value of $z$
\begin{equation}
\label{eq21} <z>_q=\int_{-\infty}^{\infty}dz\ z\  \left[\varphi^q_j(z)\right]^2=-\frac{c_q}{\omega^2}
\end{equation}
We define the auxiliary parameter
\begin{equation}
\label{eq22} \tilde p_q =\sqrt {p^2-\alpha <z>_q}
\end{equation}
and the auxiliary TDWP
\begin{equation}
\label{eq23}  \tilde U(x)=\left(m^2-\frac{1}{4}\right)\ \tan^2 x-\left (p^2-\alpha <z>_q\right)\sin^2 x
\end{equation}
We introduce the auxiliary wave function $\tilde \psi_q (x)$ which satisfies SE
\begin{equation}
\label{eq24} \tilde \psi''_q (x)+\left[\lambda_{m(q+m)}\left( \tilde p_q\right)+\frac{1}{2}-m^2-\tilde p_q^2-\tilde U(x)\right]\tilde \psi_q (x)=0
\end{equation}
Its solution is (\ref{eq11}) with taking into account the replacement $p \rightarrow \sqrt {p^2-\alpha <z>_q}$
\begin{equation}
\label{eq25} \tilde  \psi_q (x)=\cos^{1/2} x\ \bar S_{m(q+m)}\left(\sqrt {p^2-\alpha <z>_q};\sin x\right)
\end{equation}
We seek the solution of (\ref{eq16}) in the form
\begin{equation}
\label{eq26}  \Phi(x,q,z,j)\approx \varphi^q_j(z) \frac{B_{q}}{\sqrt {P_q(x)}}\cos\left(\int_0^x d\xi\ P_q(\xi) + S_q\right)
\end{equation}
We take into account the equation for $\varphi^q_j(z)$ (see argumentation in \cite{Sit20})
\begin{equation}
\label{eq27}  \left[\delta\frac{d^2}{dz^2}-\epsilon_q+\Lambda^j_q-\frac{\omega^2 z^2}{2}-c_q z\right]\varphi^q_j(z)=0
\end{equation}
Substituting (\ref{eq26}) into (\ref{eq14}) and replacing $z$ by its average value given by (\ref{eq21}) ($z \rightarrow <z>_q$) we obtain the Milne equation for $P_q(x)$
\begin{equation}
\label{eq28}  P_q^2+\frac{1}{2}\left[\frac{P_q''}{P_q}-\frac{3}{2}\frac{\left(P_q'\right)^2}{P_q^2}\right]=\epsilon_q+c_q <z>_q-\tilde U(x)
\end{equation}
We seek its approximate solution in the form
\begin{equation}
\label{eq29} P_q(x)\approx \frac{D_q}{\tilde \psi_q^2 (x)}
\end{equation}
where $D_q$ is a constant to be determined later. It is noteworthy that $P_q(x)$ from (\ref{eq29}) yields $P'_{2n}(0)=0$ because for even energy levels the wave function is symmetric ($\tilde \psi'_{2n} (0)=0$). Hence (\ref{eq4}) yields that $S_{2n}=0$ in (\ref{eq26}). Substitution of (\ref{eq29}) in (\ref{eq28}) results in the relationship
\begin{equation}
\label{eq30} \frac{D_q^2}{\tilde \psi_q^4(x)}=\lambda_{m(q+m)}\left(p\right)-\lambda_{m(q+m)}\left(\tilde p_q \right)+<z>_q\left(c_q-\alpha\right)
\end{equation}
We require that the approximate solution of SE (\ref{eq3}) (i.e., the corresponding term in (\ref{eq26})) coincides with our exact solution (\ref{eq11}) for TDWP in the crucial points $x=0$ and $x=x_{min}$. The exact wave function $\psi_q (x)$ given by (\ref{eq11}) is a normalized function in the range $-\pi/2 \leq x \leq \pi/2$ and its known values $\psi_q(0)$ and $\psi_q(x_{min})$ further replace the corresponding unknown values for the approximation (\ref{eq3}).
The requirement for (\ref{eq30}) to be satisfied at $x=0$ yields
\begin{equation}
\label{eq31} D_q=\tilde \psi_q^2(0)\sqrt{\lambda_{m(q+m)}\left(p\right)-\lambda_{m(q+m)}\left(\tilde p_q \right)+<z>_q\left(c_q-\alpha\right)}
\end{equation}
With the help of thus defined $P_{2n}(x)$ we calculate from (\ref{eq3}) (with taking into account that $P'_{2n}(0)=0$ yielding (\ref{eq8})) the wave function at $x_{min}$
\begin{equation}
\label{eq32} \psi_{2n} (x_{min})=\frac{B_{2n}}{\sqrt {P_{2n}(x_{min})}}\cos\left(\int_0^{x_{min}} d\xi\ P_{2n}(\xi)\right)
\end{equation}
As a result we obtain
\begin{equation}
\label{eq33}B_{2n}^2=\frac{\psi_{2n}^2 (x_{min})D_{2n}}{\tilde \psi_{2n}^2 (x_{min})}\ \cos^{-2}\left(D_{2n}\int_0^{x_{min}} \frac{d\xi}{\tilde \psi_{2n}^2 (\xi)}\right)
\end{equation}
Then the expression for the two-dimensional generalization of (\ref{eq9}) takes the form
\[
k(\beta, \omega)\approx\frac{1}{Z(\beta, \omega)}\Biggl\{\frac{1}{2}\sum_{n}\sum_{j=0}^{\infty}\exp\left[-\beta \Lambda^j_{2n}(\omega)\right]\frac{\psi_{2n}^2 (x_{min})D_{2n}}{\tilde \psi_{2n}^2 (x_{min})}\times
\]
\begin{equation}
\label{eq34} \cos^{-2}\left(D_{2n}\int_0^{x_{min}} \frac{d\xi}{\tilde \psi_{2n}^2 (\xi)}\right)+\sum_{l=2N+2}^{\infty}\sum_{j=0}^{\infty}\exp\left[-\beta \Lambda^j_l(\omega)\right]\Biggr\}
\end{equation}
It should be stressed that the summation over $j$ yields the same factor as that in the partition function $Z(\beta, \omega)$ and as a result they are canceled out.
Substituting (\ref{eq25}) and (\ref{eq31}) in (\ref{eq34}) we finally obtain
\[
k(\beta, \omega)\approx\left\{\sum_{q=0}^{\infty}e^{-\beta \left[\lambda_{m(q+m)}\left(p\right)-\frac{\left(c_q\right)^2}{2\omega^2}\right]}\right\}^{-1}
\Biggl \{ \frac{1}{2}\sum_{n=0}^N e^{-\beta\left[\lambda_{m(2n+m)}\left(p\right)-\frac{\left(c_{2n}\right)^2}{2\omega^2}\right]}\times
\]
\[
\cos^{-2}\Biggl\{\sqrt{\lambda_{m(2n+m)}\left(p\right)-\lambda_{m(2n+m)}\left(\sqrt{p^2-\alpha <z>_{2n}}\right)+<z>_{2n}\left(c_{2n}-\alpha\right)}\ \times
\]
\[
\bar S_{m({2n}+m)}^2\left(\sqrt {p^2-\alpha <z>_{2n}};0\right)\int_0^{x_{min}}\frac{d\xi\ \cos^{-1} \xi}{\bar S_{m({2n}+m)}^2\left(\sqrt {p^2-\alpha <z>_{2n}};\sin \xi\right)}\Biggr\}\times
\]
\[
\frac{\bar S_{m({2n}+m)}^2\left(\sqrt {p^2-\alpha <z>_{2n}};0\right)\bar S_{m({2n}+m)}^2\left(p;\sin x_{min}\right)}{\bar S_{m({2n}+m)}^2\left(\sqrt {p^2-\alpha <z>_{2n}};\sin x_{min}\right)}\times
\]
\[
\sqrt{\lambda_{m(2n+m)}\left(p\right)-\lambda_{m(2n+m)}\left(\sqrt{p^2-\alpha <z>_{2n}}\right)+<z>_{2n}\left(c_{2n}-\alpha\right)}+
\]
\begin{equation}
\label{eq35}\sum_{l=2N+2}^{\infty}e^{-\beta \left[\lambda_{m(l+m)}\left(p\right)-\frac{\left(c_{l}\right)^2}{2\omega^2}\right]}\Biggr\}
\end{equation}
where $<z>_{2n}$ is given by (\ref{eq21}). The sum over $n$ is that over the doublets below the barrier top (see the discussion below (\ref{eq2})).
\section{Results and discussion}
Notwithstanding to be cumbersome the formula (\ref{eq35}) is easily programmed in {\sl {Mathematica}} because the crucial elements $\lambda_{m(q+m)}\left(p\right)$ and $\bar S_{m(q+m)}\left(p;s\right)$ are implemented in this software package. We take the rate constant $k(\beta, \omega_{ref})$ for the internal stretching mode of the heavy atoms in the Zundel ion (with a fixed frequency $\omega_{ref}$) as a natural reference point.
For this object there are potential energy surfaces for several $R_{OO}$ as a result of quantum chemical {\it ab initio} calculations \cite{Yu16}, \cite{Xu18}. For the cases of HB in the Zundel ion with $R_{OO}=2.5\ A$, $R_{OO}=2.6\ A$ and $R_{OO}=2.7\ A$ the authors of \cite{Jan73} provide the estimates of the dimensional coupling constant $\lambda$ ($a_{21}$ in their Table.1) as 0.1 a.u., 0.1 a.u. and 0.05 a.u. respectively. For the dimensional frequency $\Omega/2$ ($a_{02}$ in their Table.1) they present the value 0.039 a.u. for all three distances. From here we obtain the reference value of $\alpha=0.6$ at $R_{OO}=2.5\ A$ and $\alpha=0.3$ at $R_{OO}=2.7\ A$. Also from the above results of \cite{Jan73} we obtain that the reference value for the dimensionless frequency of O-O stretching mode is $\omega_{ref}=1.4$. In the present article we are interested in the effect of the external vibration on the tunneling process. To make the contribution into PT rate constant from the over-barrier transition to be negligible compared with the tunneling one even at $T=400\ K$ ($\beta=0.0345$) we restrict ourselves by the case of PT for HB in the Zundel ion with very high barrier. For this reason we consider $R_{OO}=3.0\ A$ from data of \cite{Xu18}. TDWP for this case is depicted Fig.1. We carry out the parametric analysis of PT rate constant for HB in the Zundel ion with large $R_{OO}=3.0\ A$ taking the above mentioned reference value $\alpha=0.3$ and varying $\omega$ in the range $0.005 \leq \omega \leq \omega_{ref}$. In this case there are three doublets below the barrier top (see Fig.1) that means $N=2$ in (\ref{eq35}). In the calculations of the rate constant we also take into account two levels above the barrier top (i.e., replace $\infty$ in (\ref{eq35}) by $N_{max}=7$) to make sure that the contribution of the over-barrier transition can be discarded.

Fig.2 shows that at decreasing the frequency from the reference value $\omega_{ref}$ we obtain the monotonic increase of PT rate constant in agreement with the conclusion of \cite{Shi11}.
\begin{figure}
\begin{center}
\includegraphics* [width=\textwidth] {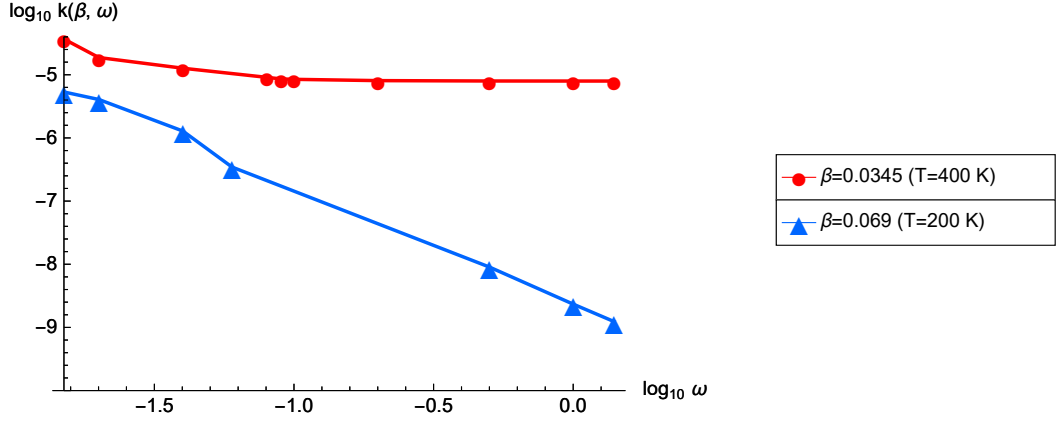}
\end{center}
\caption{The dependence of proton transfer rate constant on the frequency of the external oscillator above the critical value $\omega > \omega_c$ in the Zundel ion ${\rm{H_5O_2^{+}}}$ (oxonium hydrate) with $R_{OO}=3.0\ A$. The critical frequency is $\omega_c \approx 0.01125$ for the value of the coupling constant $\alpha=0.3$ between the proton coordinate and that of the oscillator.} \label{Fig.2}
\end{figure}
However there is a critical value of the frequency $\omega_c$ (for $\alpha=0.3$ this value is $\omega_c \approx 0.01125$) below which a drastic change of the behavior takes place. In Fig.3, Fig.4 and Fig.5 the dependence of PT rate constant on the frequency of the oscillator at $\omega<\omega_c$ is depicted.
\begin{figure}
\begin{center}
\includegraphics* [width=\textwidth] {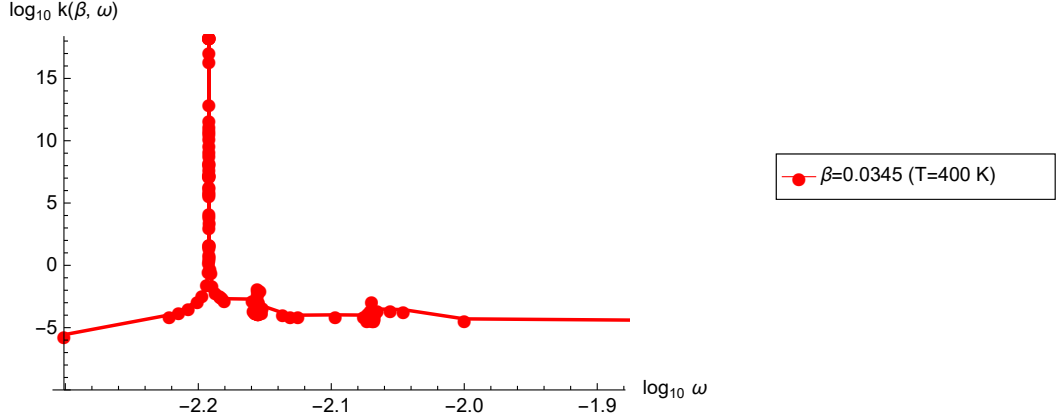}
\end{center}
\caption{The dependence of proton transfer rate constant on the frequency of the external oscillator below the critical value $\omega < \omega_c$ in the Zundel ion ${\rm{H_5O_2^{+}}}$ (oxonium hydrate) with $R_{OO}=3.0\ A$ at high temperature ($T=400\ K$ ($\beta=0.0345$). The critical frequency is $\omega_c \approx 0.01125$ for the value of the coupling constant $\alpha=0.3$ between the proton coordinate and that of the oscillator. Low resolution picture of the whole interval $0.005 < \omega < \omega_c$.} \label{Fig.3}
\end{figure}
\begin{figure}
\begin{center}
\includegraphics* [width=\textwidth] {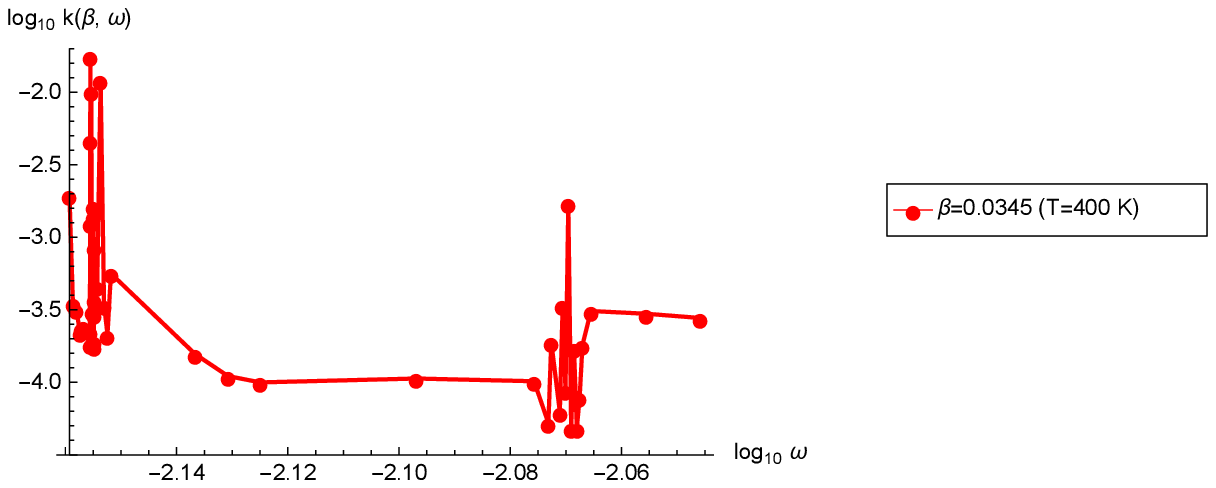}
\end{center}
\caption{The dependence of proton transfer rate constant on the frequency of the external oscillator below the critical value $\omega < \omega_c$ in the Zundel ion ${\rm{H_5O_2^{+}}}$ (oxonium hydrate) with $R_{OO}=3.0\ A$ at high temperature ($T=400\ K$ ($\beta=0.0345$). The critical frequency is $\omega_c \approx 0.01125$ for the value of the coupling constant $\alpha=0.3$ between the proton coordinate and that of the oscillator. High resolution picture of the interval $0.00693 < \omega < 0.009$.} \label{Fig.4}
\end{figure}
\begin{figure}
\begin{center}
\includegraphics* [width=\textwidth] {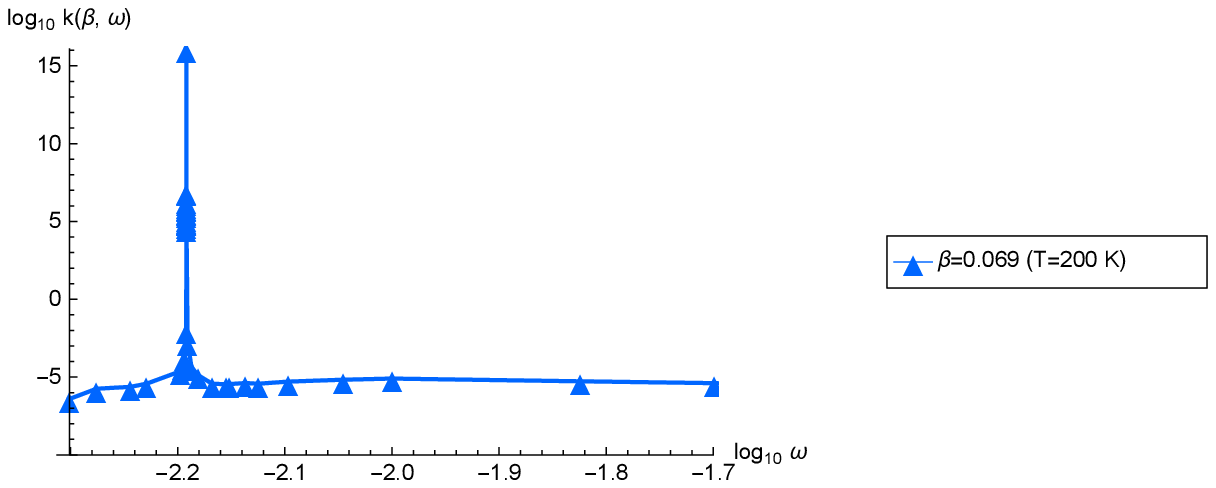}
\end{center}
\caption{The dependence of proton transfer rate constant on the frequency of the external oscillator below the critical value $\omega < \omega_c$ in the Zundel ion ${\rm{H_5O_2^{+}}}$ (oxonium hydrate) with $R_{OO}=3.0\ A$ at low temperature ($T=200\ K$ ($\beta=0.069$). The critical frequency is $\omega_c \approx 0.01125$ for the value of the coupling constant $\alpha=0.3$ between the proton coordinate and that of the oscillator.} \label{Fig.5}
\end{figure}
For a given value of the coupling constant $\alpha$ at the corresponding $\omega_c$ there is $\Lambda^0_{1}=\Lambda^0_{0}$ and a reversal in the total energy levels takes place (the energy levels of TDWP do not depend on $\omega$ and retain their natural order $\epsilon_{q+1}>\epsilon_q$). At $\omega > \omega_c$ we have the normal sequence $\Lambda^0_{2n+1} > \Lambda^0_{2n}$ where $n=0,1,2,...$ while at $\omega < \omega_c$ an anomalous picture $\Lambda^0_{2n+1} < \Lambda^0_{2n}$ occurs at first for the ground state doublet ($n=0$) and at further decrease of the frequency for higher ones below the barrier top (see, e.g., the case for $\omega_{max}$ in Fig.1). This transformation leads to an extraordinary alteration in the behavior of PT rate constant. Fig.3, Fig.4 and Fig.5 vividly exhibit that in this case there are very rich manifestations of resonant activation, i.e., the bell-shaped peaks of PT rate enhancement by the external vibration at its symmetric coupling to the proton coordinate. The height of the main peak at $\omega_{max}=0.006429109800232905$ is temperature dependent (e.g., $k(0.046,\ \omega_{max})/k(0.046,\ \omega_{ref})=5.31 \cdot 10^{22}$ at $T=300 K$ and $k(0.0345,\ \omega_{max})/k(0.0345,\ \omega_{ref})=3.13 \cdot 10^{23}$ at $T=400 K$). Fig.3 shows that at high temperature $\beta=0.0345$ the approach to the main peak from the side of higher frequencies is not smooth. There is a sequence of comb-like regions of increasing intensity with the decrease of the frequency $\omega$ (see Fig.4 for higher resolution picture). The intensity of these combs decreases with the decrease of temperature and at $\beta=0.069$ they are not discernable (see Fig.5). At $\alpha=0.3$ and $\omega<\omega_c$ the ground state doublet approaches the bottom of TDWP (see, e.g., Fig.1) and at $\omega < 0.005$ the former becomes below the latter.

By attaining the resonance condition $\omega_{max}$ one can obtain a very efficient mechanism of PT rate enhancement. For $\alpha=0.3$ we have the acceleration up to 23 orders of magnitude compared with the reference value $\omega_{ref}=1.4$. The mathematical reason for the phenomenon of such PT resonance activation is in the fact that the function $\bar S_{m({2n}+m)}\left(\sqrt {p^2-\alpha \left(-\frac{c_{2n}}{\omega^2}\right)};\sin x_{min}\right)$ taking place in the denominators of (\ref{eq35}) becomes extremely small for the second doublet $n=1$ at $\omega=\omega_{max}$. In our opinion the descriptive physical origin of the phenomenon can be revealed from the following empirical observation. Let us consider the wave functions in the left and the right wells for the $j$-th doublet ($j=1,2,3$) defined as usual $\psi_R^{(j)}=1/\sqrt 2\left(\psi_+^{(j)}-\psi_-^{(j)}\right)$ and $\psi_L^{(j)}=1/\sqrt 2\left(\psi_+^{(j)}+\psi_-^{(j)}\right)$ respectively where $\psi_{\pm}^{(j)}$ are given by (\ref{eq11}). Here $+$ means the upper energy level in the doublet while $-$ means the lower one. Then we recall the notion of the Rabi frequency in energetic units (multiplied by the Planck constant) as the module of the interaction energy, i.e., that of the product of the electromagnetic field strength and the matrix element of the dipole moment for the transition between the corresponding energy levels. The dimensional resonance condition $E_f-E_i = \hbar \Omega_{if}^{Rabi}=\mid H_{int} \mid$ in the dimensionless form is $\epsilon_f-\epsilon_i = \sqrt {2\delta}\ \omega_{if}^{Rabi}=\mid h_{int} \mid$. Analogously we equate the difference between the energy levels $\epsilon_+^{(j)}-\epsilon_{-}^{(j)}$ in the $j$-th doublet and the module of the matrix element of the interaction energy term $\mid \alpha z\sin^2 x \mid$ from (\ref{eq14}) with the functions $\psi_R^{(j)}$ and $\psi_L^{(j)}$. For $\sin^2 x$ we take the value of its matrix element
\begin{equation}
\label{eq36} <\psi_R^{(j)} \mid \sin^2 x \mid \psi_L^{(j)}>=\int_{-\pi/2}^{\pi/2} dx\ \psi_R^{(j)}\sin^2 x\ \psi_L^{(j)}
\end{equation}
For $z$ we take the average $<z>_{-}$ given by (\ref{eq21}), i.e., $ <z>_{-} = -c_{-}^{(j)}/ \omega^2$. As a result we have an empirical relationship
\begin{equation}
\label{eq37}\epsilon_+^{(j)}-\epsilon_{-}^{(j)}=\alpha\ \frac{\mid -c_{-}^{(j)} <\psi_R^{(j)} \mid \sin^2 x \mid \psi_L^{(j)}>\mid}{\omega^2}
\end{equation}
From (\ref{eq37}) we obtain the resonance frequency $\omega_{\pm}^{(j)}$
\begin{equation}
\label{eq38}\omega_{\pm}^{(j)}=\sqrt {\frac{\alpha \mid -c_{-}^{(j)} <\psi_R^{(j)} \mid \sin^2 x \mid \psi_L^{(j)}>\mid }{\epsilon_+^{(j)}-\epsilon_{-}^{(j)}
}}
\end{equation}
At $\alpha=0.3$ and $\delta=1/8$ we have for $j=2$, i.e., for the second doublet $\omega_{\pm}^{(2)}=0.00673$ which is rather close to $\omega_{max}\approx 0.00643$ for the main peak. In our opinion such quantitative agreement can not be fortuitous. It suggests the physical interpretation of PT resonant activation as an analog of the Rabi transition between the left and the right wells under the influence of the vibration with a suitable frequency applied to the proton in DWP. The case $j=1$ yields the resonance frequency $\omega_{\pm}^{(1)}=0.00795$ which is within the range of the right comb-like region in Fig.4. Constructing various matrix elements between wave functions of different doublets for both wells $\psi_R^{in}=1/\sqrt 2\left(\psi_i-\psi_n\right)$ and $\psi_L^{lm}=1/\sqrt 2\left(\psi_l+\psi_m\right)$ yields
\begin{equation}
\label{eq39}\epsilon_+^{(j)}-\epsilon_{-}^{(j)}=\alpha\ \frac{\mid -c_k <\psi_R^{in} \mid \sin^2 x \mid \psi_L^{lm}>\mid}{\omega^2}
\end{equation}
where $j=1,2,3$ and $\left\{k,i,n,l,m=0,1,2,3,4,5\right\}$. Then we obtain for the third doublet $j=3$ the resonance frequencies: $0.00722$ at\\ $\left\{l=2,m=5,i=1,k=n=0\right\}$; $0.00725$ at $\left\{k=l=0,m=3,i=1,n=4\right\}$; $0.00753$ at\\
$\left\{l=1,k=m=4,i=2,n=5\right\}$ which are within the range of the left comb-like region in Fig.4. Constructing various matrix elements between wave functions of different doublets for the left well yields
\begin{equation}
\label{eq40}\epsilon_+^{(j)}-\epsilon_{-}^{(j)}=\alpha\ \frac{\mid -c_k <\psi_L^{lm} \mid \sin^2 x \mid \psi_L^{l'm'}>\mid}{\omega^2}
\end{equation}
Then we obtain for the third doublet $j=3$ the resonance frequencies:
$0.00787$ at\\
$\left\{l=1,k=m=4,l'=5,m'=4\right\}$ which is within the range of the right comb-like region in Fig.4 and
$0.00723$ at $\left\{l=0,m=3,l'=4,k=m'=5\right\}$ which is within the range of the left comb-like region in Fig.4. In our opinion these numerous  resonance frequencies provide qualitative explanation of severe oscillations in Fig.4.

Also in this connection it is worthy to note that for the symmetric mode coupling ($H_{int}=\lambda Z X^2$) the effect of resonant activation results from the term $-\left(c_q\right)^2/2\omega^2$ in the total energy levels $\Lambda^k_q$ (\ref{eq19}). The energy levels of TDWP $\epsilon_q$ are re-normalized due to the coupling of the proton coordinate to the oscillator. For anti-symmetric ($H_{int}=\lambda Z X$) and squeezed ($H_{int}=\lambda Z^2 X^2$) mode couplings this effect is absent. In the former case the coupling strength is zero $c^{\{as\}}_q=0$ due to the symmetry of the wave functions \cite{Sit20} that leads to the actual lack of the crucial term $-\left(c^{\{as\}}_q\right)^2/\left(2\omega^2\right)$ in the formula (\ref{eq19}) for $\Lambda^k_q$. In the latter case the expression for $\Lambda^k_q$  \cite{Sit20}
\begin{equation}
\label{eq41} \Lambda^k_q|^{\{sq\}}\approx\lambda_{m(q+m)}\left(p\right)+\frac{1}{2}-m^2-p^2+(4k+1)\sqrt{\frac{\delta\left(\omega^2+2c^{\{sq\}}_q\right)}{2}}
\end{equation}
does not contain the required term at all. For instance the interaction of the proton in HB with an IR laser field in the dipole approximation (the dipole moment $d \propto x$) belongs to the anti-symmetric type and does not fit our requirement for PT resonant activation by a low-frequency vibration (high-frequency Rabi transitions between different doublets stimulated by an IR laser field certainly can considerably interfere PT process). Only taking into account that a realistic dipole moment contains the appropriate higher order contributions ($d \propto x+const\ x^2+...$) may provide the required type of interaction in this case.

We conclude that the suggested approach enables one to obtain an analytically tractable expression for proton transfer rate constant in a hydrogen bond. It is based on the Schr\"odinger equation with the model Hamiltonian taking into account only the proton coordinate and an external oscillator coupled to it (the heavy atoms stretching mode, a low-frequency vibration of the protein scaffold in an enzyme, etc). The literature data from quantum chemical {\it ab initio} calculations of the potential energy surface are transformed into the parameters of the model trigonometric double-well potential. For the two-dimensional Schr\"odinger equation with this potential the analytic solution within the framework of the standard adiabatic approximation is available. The parameters of the model for the Zundel ion in the case of the heavy atoms stretching mode are extracted from the literature data on IR spectroscopy and serve as a reference point. The approach yields the pronounced resonant effect of proton transfer acceleration in some frequency range of the oscillator (below the corresponding critical value of the frequency) at its symmetric coupling to the proton coordinate. The phenomenon is absent for anti-symmetric and squeezed mode couplings.
\section{Appendix}
In dimensional units the one-dimensional SE for a quantum particle with the reduced mass $M$ (proton in our case of usual HB or deuterium in the case of a deuterated HB) has the form
\begin{equation}
\label{eq42} \frac{d^2 \psi (X)}{dX^2}+\frac{2M}{\hbar^2}\left[E-V(X)\right]\psi (X)=0
\end{equation}
where $-L \leq X \leq L$ and $V(X)$ is a DWP. The latter is assumed to be infinite at the boundaries of the finite interval for the spatial variable $X=\pm L$.
The dimensionless values for the distance $x$, the potential $U(x)$ and the energy $\epsilon$ are introduced as follows
\begin{equation}
\label{eq43} x=\frac{\pi X}{2L};\ \ \ \ \ \ \ \ \ \ \ \ \ \ \ \ \ U(x)=\frac{8ML^2}{\hbar^2 \pi^2}V(X);\ \ \ \ \ \ \ \ \ \ \ \ \ \epsilon=\frac{8ML^2E}{\hbar^2 \pi^2}
\end{equation}
where $-\pi/2\leq x \leq \pi/2$. As a result we obtain the dimensionless SE (\ref{eq1}). In the case of the trigonometric DWP (\ref{eq10})
the transformation formulas for the parameters $\{m,p\}$ into $\{B,D\}$ ($B$ is the barrier height and $D$ is the barrier width) are \cite{Sit19}
\begin{equation}
\label{eq44}p=\frac{\sqrt {B}}{1-\left[\cos\left(D/2\right)\right]^2}; \ \ \ \ \ \ \ \ \ \ \ \ \ \ \ \ \  m^2-\frac{1}{4}=\frac{B\left[\cos\left(D/2\right)\right]^4}{\left\{1-\left[\cos\left(D/2\right)\right]^2\right\}^2}
\end{equation}
The Hamiltonian of the two-dimensional SE includes the spatial variable $Z$ (e.g., that for the reduced mass of the heavy atoms in HB which in the case of the Zundel ion is the O-O stretching mode) of the harmonic potential $\Omega Z^2/2$ with the frequency $\Omega$. We introduce the dimensionless distance $x=\pi X/\left(2L\right)$ where $-\pi/2\leq x \leq \pi/2$ and dimensionless coordinate $z=\pi Z/\left(2L\right)$ where $-\infty < z < \infty$. The dimensionless coupling constant $\alpha$ in (\ref{eq14}) for the symmetric mode coupling (this case was proved in \cite{Jan73} to be pertinent for the Zundel ion), the dimensionless inverse temperature $\beta$ in (\ref{eq20}), (\ref{eq34}) and (\ref{eq35}) and the dimensionless frequency $\omega$ in (\ref{eq14}) are
\begin{equation}
\label{eq45}
\alpha=\frac{2^6\lambda ML^5}{\hbar^2 \pi^5};\ \ \beta=\frac{\hbar^2\pi^2}{8ML^2k_BT};\ \ \omega=\frac{4\sqrt{2M\mu}L^2\Omega}{\hbar \pi^2};\ \ \mu=\frac{M_1M_2}{M_1+M_2}
\end{equation}
Here $\lambda$ is a dimensional coupling constant for the case of the symmetric mode coupling term ($\lambda Z X^2$) and $\mu$ is the reduced mass of the heavy atoms in HB $A_1-H \cdot\cdot\cdot A_2$. In the case of the Zundel ion it is $\mu=M_O/2$. As a result $\delta$ in (\ref{eq14}) is $\delta=M/\mu=2M/M_O$. Taking the proton mass $M=1\ a.u$ and that of the oxygen atom $M_O=16\ a.u.$ we have $\delta=1/8$.

Acknowledgements. The author is grateful to Prof. Yu.F. Zuev for helpful discussions. The work was supported from the government assignment for FRC Kazan Scientific Center of RAS.\\

\end{document}